\documentclass[10pt, conference, compsocconf]{IEEEtran}

\usepackage{cite}
\usepackage{hyperref}
\hypersetup{
	colorlinks=false,
	pdfborder={0 0 0},
}
\usepackage[pdftex]{graphicx}
\usepackage{url}
\begin{document}

\title{GHTraffic: A Dataset for Reproducible Research in Service-Oriented Computing}


\author{\IEEEauthorblockN{Thilini Bhagya, Jens Dietrich, Hans Guesgen}
\IEEEauthorblockA{Massey University\\
Palmerston North, New Zealand\\
Email: \{t.bhagya, j.b.dietrich, h.w.guesgen\}@massey.ac.nz}
\and
\IEEEauthorblockN{Steve Versteeg}
\IEEEauthorblockA{CA Technologies\\
	Melbourne, Australia\\
	Email: steve.versteeg@ca.com}
}
\maketitle

\begin{abstract}
We present GHTraffic, a dataset of significant size comprising HTTP transactions extracted from GitHub data and augmented with synthetic transaction data. The dataset facilitates reproducible research on many aspects of service-oriented computing. This paper discusses use cases for such a dataset and extracts a set of requirements from these use cases. We then discuss the design of GHTraffic, and the methods and tool used to construct it. We conclude our contribution with some selective metrics that characterise GHTraffic.

\end{abstract}


\IEEEpeerreviewmaketitle

% body 
\section{Introduction}
Service-Oriented Computing (SOC) is a popular approach to facilitate the development of large, modular applications using diverse technologies. There is a range of technologies that have been used in SOC, with early attempts to establish standards around the SOAP~\cite{box1999soap} and WSDL~\cite{christensen2001web} protocols. In recent years,  \textit{RESTful services}~\cite{fielding2000architectural}, a more lightweight approach closely aligned with the Hypertext Transfer Protocol (HTTP)~\cite{fielding1999hypertext}, have become mainstream. 

When using HTTP-based services, different parts of the application cooperate by sending and responding to HTTP requests, typically in order to access and manipulate resources. The ubiquitousness of the HTTP means that clients and servers can be easily implemented in a wide range of languages and deployed on many platforms. While this is useful in itself to architect and design large applications, this approach is now increasingly used to facilitate the development of product ecosystems around successful services. Examples include the APIs that can be used to access the services of Google, Facebook, Amazon, and Netflix.

This has created new challenges for both  the research and the engineering community. Of particular interest are scalability, reliability, and security of (systems using and providing) services.

Like other fields of computing research, studies of SOC should aim for reproducibility~\cite{peng2011reproducible,collberg2016repeatability}. There is a wider push for reproducibility in computing research, with some disciplines now including research artefact evaluation as part of the standard peer-review process~\cite{krishnamurthi2015real}. One way to facilitate the reproducibility and also the dissemination of research is the provision of standardised datasets. By using carefully sourced and/or constructed datasets, research results become (1) easier to reproduce (2) comparable (i.e., results from different studies can be compared), and (3) generalisable (i.e., we can assume with a certain amount of confidence that results from a study can be applied to other data/systems that were not studied).

The purpose of this paper is to provide such a dataset, GHTraffic. We extract a base dataset from a successful, large-scale service, GitHub, by reverse-engineering API interactions from existing repository snapshots. We then enrich the dataset to include API interactions that cannot be recovered from snapshots, namely (non-state-changing) queries. This results in a large, rich, and diverse dataset. We argue that this can be used for a wide range of studies, including performance benchmarking and service virtualisation.  

The rest of the paper is organised as follows. Use cases and requirements are discussed in detail in Section~\ref{sec:usecases}, followed by an overview of related work in Section~\ref{sec:relatedwork}. The construction of the dataset is discussed in Section~\ref{sec:methodology}. Section~\ref{sec:metrics} and~\ref{sec:usage} present the results of some measurements on the dataset and provide basic instructions how to obtain and use GHTraffic. We discuss threats to validity in Section~\ref{sec:threats} and Section~\ref{sec:conclusion} concludes our contribution.

\section{Use Cases and Requirements} \label{sec:usecases}

\subsection{Performance Benchmarking} 
Modern enterprise applications usually cooperate with a variety of software services such as Web servers, application servers, databases, proxies, and Web service clients to perform their functionalities. These services need to be tested in order to ensure that they are able to deal with large data and transaction volumes. In particular, performance benchmarking can provide useful indications about how services behave under different load conditions. A typical benchmarking tool generates synthetic workloads or replays recorded real-world network traffic in order to simulate realistic workloads and measures performance-related metrics, such as latency and throughput. 

A dataset that is large, complex, and extracted from actual network traffic facilitates the benchmarking of such systems with non-trivial, realistic workloads. 

\subsection{Functional Testing}
A standard dataset can also be employed for functional testing. For instance, it could be used to test a generic REST framework with a CRUD back-end provided by a (non-SQL) database. This would take advantage of the fact that such a dataset encodes a certain semantics, usually a combination of the standard HTTP semantics (for instance, the idempotency of certain methods) plus additional, application-specific rules and constraints. In other words, a suitable dataset can provide an \textit{oracle} of correct system and service behaviour. As an example, consider an HTTP GET request to a named resource. This request should result in a response with 200 status code if there was an earlier successful POST transaction for the resource and no successful DELETE transaction between the POST and the GET, and 404 otherwise. A suitable dataset should contain transaction sequences to reflect such behavioural patterns. 

\subsection{Service Virtualisation}
Service Virtualisation (SV)\cite{Michelsen:2012:SVR:2385447} is an approach to build a semantic model of a service based on recorded traffic. For instance, SV will try to simulate the behaviour of an actual service by generating responses using models inferred from recorded transactions. This inference is usually done by means of supervised machine learning. The main application is to test systems that heavily rely on external \textit{black-box} services in isolation. This corresponds to (automatically created) mock objects popular in application testing~\cite{mackinnon2000endo}. 

A suitable standardised dataset could be used to test SV. It would provide an oracle of actual service behaviour to be used in order to assess the quality of inferred behaviour.

\subsection{Requirements}
From the use cases above, we extract the following set of requirements to guide the construction of GHTraffic.

\begin{itemize}
	
	\item[R1] \textbf{Large, yet manageable:} a good dataset should be of significant size to facilitate the use cases outlined and obtain results that are generalisable. However, this often conflicts with usability as experiments on large datasets are more difficult to set up and time-consuming. This can be addressed by providing several editions of different sizes.
	
	\item[R2] \textbf{Ease of use:} a good dataset should be presented in a format that is easy to process and preferably includes scripts to facilitate the processing and analysis of data, and a schema that (formally) describes the format used to represent data.
	
	\item[R3] \textbf{Reproducible, independent, and derived from principles}: a good dataset should not be produced ad-hoc, but extracted from real-world data or synthesised using a well-defined process unbiased by its use for one particular experiment. 
	
	\item[R4] \textbf{Current:} a good dataset should reflect the state-of-the-art use of HTTP-based services. While this is difficult to assess in general, we argue that by extracting the dataset from the traffic of one of the most successful active Web services known for its excellent scalability and robustness, this can be achieved. 
	
	\item[R5] \textbf{Precise and following standards:} a good dataset should contain transactions that comply with the syntax and semantics of HTTP, and the service(s) used. 
	
	\item[R6] \textbf{Diverse:} a good dataset should support a wide set of HTTP features, such as various HTTP methods and status codes. In particular, it should go beyond the exclusive use of POST and GET requests which is a characteristic of older-generation Web applications designed for browser-based clients. 
	
\end{itemize}

\section{Related Work} \label{sec:relatedwork}

%This section primarily reviews the standard benchmarks and datasets containing HTTP message traces.

\textbf{SPECweb2009}~\cite{specweb2009} is a standardised Web server benchmark produced by the Standard Performance Evaluation Corporation (SPEC). It is designed to evaluate a Web server ability to serve static and dynamic page requests. The benchmark comprises four distinct HTTP workloads to simulate three common types of consumer activities. The first workload is based on online banking, the second one is based on an e-commerce application, and the third one uses a scenario where support patches for computer applications are downloaded. All these workloads were developed by analysing log files of several popular Internet servers. The benchmark uses one or more client systems to generate HTTP workload for the server according to the specified workload characteristics. Each client sends HTTP requests to the server and then validates the response received. However, SPECweb2009 uses only HTTP 1.1 GET and POST requests and all of these requests are expected to result in responses with a 200 status code. In particular, server errors are communicated back to clients by generating error pages that return 200.

\textbf{TPC Benchmark W} (TPC-W)~\cite{smith2000tpc} from the Transaction Processing Council is a notable open-source Web benchmark specifically targeted at measuring the performance of e-commerce systems. TPC-W simulates the principal transaction types of a retail store that sells products over the Internet. The workload of this benchmark specifies emulated browsers that generate Web interactions which represent typical browsing, searching, and ordering activities. It creates different GET and POST requests for specific documents and collects performance data. All these requests are expected to result in responses with 200 status code. 

\textbf{Rice University Bidding System (RUBiS)}~\cite{ow22013rubis} is another open-source Web benchmark. It is based on an online auction site, modelled after eBay. This benchmark implements the core functionality of an auction site, in particular, selling, browsing, and bidding. The benchmark workload relies on a number of browser emulators that mimic the basic network interactions of real Web browsers. Read and write interactions are implemented using HTTP GET and POST requests. 

\textbf{DARPA dataset}~\cite{darpadataset}  by the MIT Lincoln Laboratory is a widely used evaluation dataset in intrusion detection research. There were three major releases. Each release contains tcpdump files carrying a wide variety of normal and malicious Web traffic simulated in a military network environment. These network packet dumps can be used as a direct input to packet filtering engines like Wireshark to extract sub-datasets which contain only HTTP request/response message traces as relevant to our work. In particular, an HTTP dataset which comprises 25,000 transactions can be obtained from DARPA 2000 tcpdump files. All these transactions used HTTP 1.0 GET requests and returned 200. 

\textbf{CSIC 2010}~\cite{gimnez2010http} by the Information Security Institute of Spanish Research National Council is another publicly available dataset, designed for the purpose of testing intrusion detection systems. It contains normal and anomalous HTTP 1.1 POST and GET requests targeting an e-commerce Web application. However, the dataset does not contain response data. 

Another example of an HTTP message traces dataset is described in the work of Versteeg et al.~\cite{versteeg2016opaque,versteeg2017entropy}. The authors used a relatively small dataset to study \textbf{Opaque SV}. This dataset consists of 1,825 request/response message traces collected through the Twitter REST API\footnote{\url{https://developer.twitter.com/en/docs} [accessed Feb. 02 2018]}. It contains both POST and GET requests which return 200. However, the dataset does not cover all the HTTP methods such as PUT and DELETE. Besides, it is not publicly available for research purposes due to Twitter's terms of service.

Table~\ref{tab:relatedwork:overview} summarises related benchmarks and datasets showing their request types, response codes, and transaction count. It is apparent that all these datasets only use a small fraction of the HTTP in terms of methods and status codes. They are somehow biased towards performance testing for older Web server where (static) pages are retrieved and in some cases created. They do not reflect the richness of modern Web APIs that take advantage of a much larger part of the HTTP. 

Standard datasets have been widely used to support research in many other areas of computer science. For instance, the programming language and software engineering communities use datasets such as \textbf{DaCapo}~\cite{blackburn2006dacapo} and \textbf{Qualitas Corpus}/\textbf{XCorpus}~\cite{tempero2010qualitas,dietrich2017xcorpus} for benchmarking and empirical studies on source code. \textbf{Sourcerer}~\cite{bajracharya2006sourcerer} is an infrastructure for large-scale collection and analysis of open source code. The Sourcerer database is populated with more than 2000 real-world open source projects taken from Sourceforge, Apache, and Java.net. 

The machine learning community uses several standardised datasets. This includes \textbf{UCI Machine Learning Repository} \cite{frank2010uci} by the Center for Machine Learning and Intelligent Systems at the University of California, Irvine. It provides a collection of benchmark datasets which can be used for the empirical analysis of learning algorithms. Another example is \textbf{Kaggle}\footnote{\url{https://www.kaggle.com/datasets/} [accessed Feb. 02 2018]}. 

\begin{table}[!t]
	\renewcommand{\arraystretch}{1.3}
	\caption{Overview of HTTP Benchmarks and Datasets}
	\label{tab:relatedwork:overview}
	\centering
	\footnotesize
	\begin{tabular}{|c|c|c|c|}
		\hline
		Name & HTTP Method     & Response Code & Count\\    
		\hline
		TPC-W & GET, POST & 200 & 13,500,000\\
		\hline
		RUBiS & GET, POST & 200 &4,030,000 \\ 
		\hline
		DARPA 2000 & GET & 200 & 25,000\\ 
		\hline
		CSIC 2010 & GET, POST & - & 36,000\\
		\hline    
		Opaque SV & GET, POST & 200 & 1,825\\
		\hline
	\end{tabular}
\end{table}

\section{Methodology} \label{sec:methodology}

\subsection{Input Data Selection}
Over the past few years, GitHub\footnote{\url{https://github.com/} [accessed Feb. 02 2018]} has emerged as the dominant platform for collaborative software engineering. It contains a rich set of features to manage code-related artefacts, including commits, pull requests, and issues. 

There are several clients provided by GitHub that can be used to access its services, including the Web front-end and the desktop app. Many developers also use the standard git command line interface (CLI).  In order to facilitate the development of a rich product ecosystem to access its services, GitHub also provides a REST API\footnote{\url{https://developer.github.com/v3/} [accessed Feb. 02 2018]}. This allows third parties to integrate GitHub services into their products. Examples include mobile clients as well as IDE and build tool integrations (plugins).

The GitHub REST API provides a rich set of services to create, read, update, and delete resources related to the core GitHub functionality. It employs a large subset of HTTP features for this purpose and is therefore semantically richer than the datasets discussed on Section~\ref{sec:relatedwork}. Unfortunately, GitHub does not provide direct access to the recorded API interactions, so this information cannot be directly used for dataset construction.

An interesting use of the GitHub REST API for research purposes is GHTorrent~\cite{gousios2013ghtorrent}. This project uses the API to harvest information from repositories and stores that information by creating snapshots. These snapshots can then be downloaded and imported into a local MongoDB or MySQL database and queried. As of Feb. 02 2018, GHTorrent offers more than fifteen terabytes of downloadable snapshots. These snapshots have already been used in empirical studies, examples include Gousios et al. work on the pull-based software development model~\cite{gousios2014exploratory} and Vasilescu et al. work on the use of crowd-sourced knowledge in software development~\cite{vasilescu2013stackoverflow}.

While GHTorrent provides a static view on the state of GitHub at certain points in time, we are interested in a more dynamic view of how interactions of clients with the repository have created this state. The basic idea is to reverse-engineer the respective API interactions (i.e., HTTP transactions) by cross-referencing GHTorrent data with GitHub API functions. This has some obvious limitations. Firstly, we do not know whether all of these records were created via the REST API. They could have been created or altered using a different, or older version of the API, or via GitHub internal systems that bypass the API. We do not consider this as a significant limitation. As far as the data inferred transactions are concerned, this will only have an impact on the user-agent header. Secondly, the static data of the snapshots means that certain API interactions are not visible. This includes all read access (i.e., GET requests), requests that fail (e.g., a DELETE request resulting in a 404 response code will have no effect on the database), and shadowed requests (e.g., a successful PUT request followed by a successful DELETE request). To deal with those un-observable requests, we decided to augment the dataset with synthetic data. 

\subsection{Scope}
\begin{figure}[!t]
	\centering
	\includegraphics[width=\columnwidth]{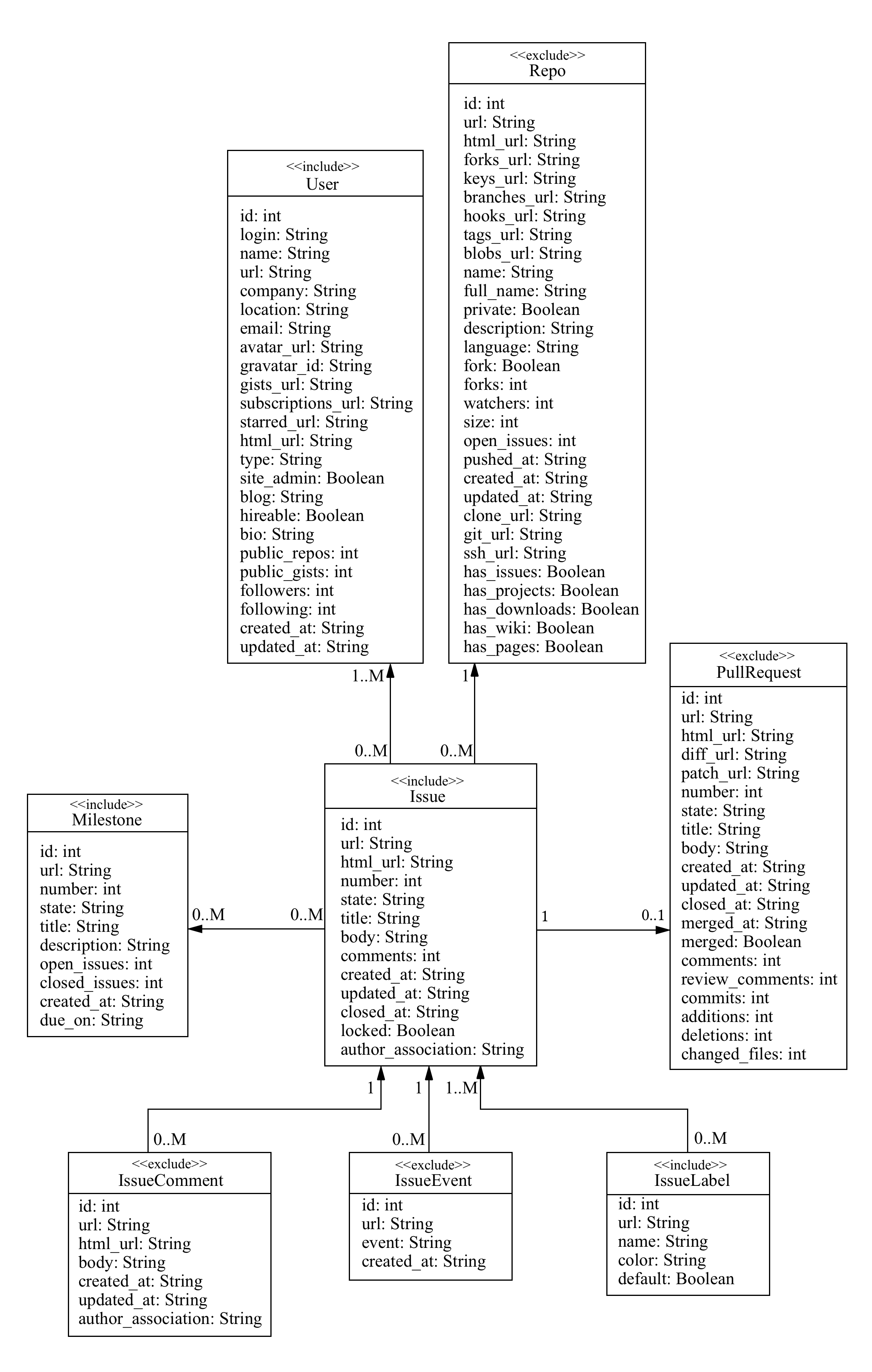}
	\caption{GitHub's data schema (UML 2.0). The stereotypes indicate which entities were included in the construction of GHTraffic dataset.}
	\label{fig:dataschema}
\end{figure}

GHTorrent collects a large amount of data on the terabyte scale. To make the data volume more manageable (R1), we decided to focus on a particular subset of GHTorrent, the issue tracking system. The issue tracking system itself references other entities\footnote{Entity is used in this paragraph in the context of entity-relationship data modelling~\cite{chen1976entity}, as opposed to the use of entity in the context of HTTP as defined by\cite[Sect. 7]{fielding1999hypertext}} of the overall data model. The respective model is depicted in Figure~\ref{fig:dataschema}. It is a refined version of the relational schema used in GHTorrent\footnote{\url{http://ghtorrent.org/relational.html} [accessed Feb. 02 2018]}.

Issues reference multiple other entities such as comments, milestones, labels, and users. While it is important to model some of them to facilitate our use cases, we decided to limit this to user, milestone, and label data. In particular, while issue comments look like integral parts of the issue tracking system, they are modelled in a relational style as one-to-many relationships via back-references. This means that comments reference the issue they are associated with, but issues do not directly reference comments\footnote{The JSON representation of an issue contains a field \texttt{comments}, but this contains only the number of comments for the respective issue. This number can then be used to construct comments queries.}. 

The design of GHTraffic is driven by the use cases and the requirements derived from them. We wanted to construct a dataset that is large and diverse%complex
, and uses the features seen in modern Web services. %in particular, the use of multiple HTTP methods. 
This can be achieved by restricting the dataset to user, milestone, and label. Adding issue comments and other related data does increase the size further but does not add new features to the dataset. On the other hand, the increased size makes the dataset less manageable. As we will demonstrate in Section~\ref{sec:metrics}, the dataset is already sufficiently large. 

Data represented in different entities is usually inlined in data returned by API calls. This means that if issue information is returned via the API, the JSON representation of the issue contains information about the issue and a summary of the users, labels, and milestones associated with it. Part of this information are URLs that can be used to query the full information for the respective entity. We treat these URLs as external, un-resolved references in the sense that our dataset does contain transactions to create, modify, delete or query these resources. Note that the GitHub API already uses references to external resources for which resolution cannot be guaranteed, an example for this is the \texttt{gravatar\_id} attribute pointing to a picture of the user provided by the gravatar\footnote{\url{https://pt.gravatar.com/} [accessed Feb. 02 2018]} service. 

The GHTraffic dataset is based on the Aug. 04 2015 GHTorrent snapshot\footnote{The respective dump is available from \url{http://ghtorrent-downloads.ewi.tudelft.nl/mongo-full/issues-dump.2015-08-04.tar.gz} [accessed Feb. 02 2018]. The download size is 6,128 MB which results in a 48.29 GB database with 21,077,018 records after restoring.}. This is the largest release of issues MongoDB database dumps as of Feb. 02 2018.

\subsection{Processing {Pipeline}} \label{ssec:pipeline}
\begin{figure}[!t]
	\centering
	\includegraphics[width=\columnwidth]{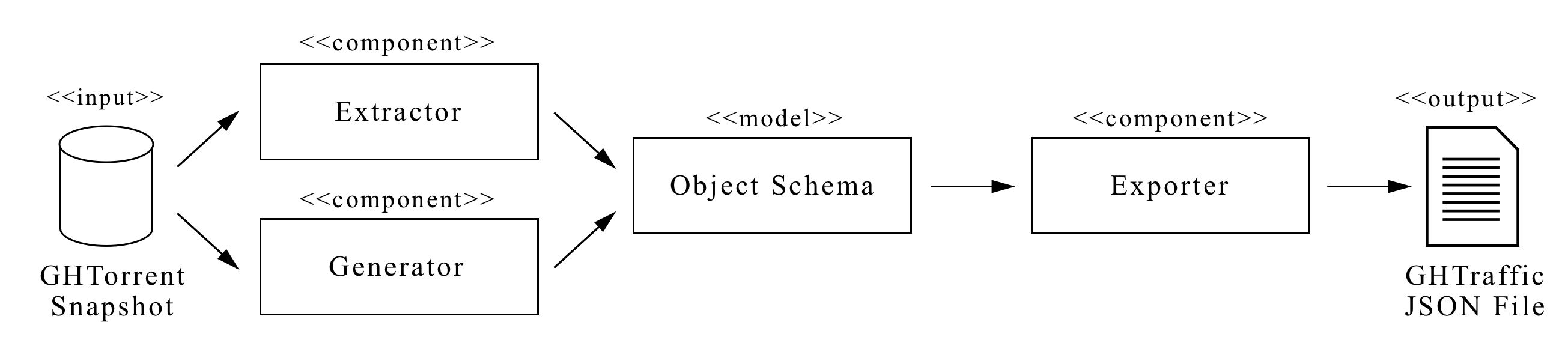}
	\caption{The processing pipeline}
	\label{fig:pipeline}
\end{figure}

An abstract overview of the infrastructure used to create the GHTraffic dataset is shown in Figure \ref{fig:pipeline}. GHTorrent snapshots are accessed by two core components, the \textit{Extractor} and the \textit{Generator}, the purpose of both is to create HTTP transactions. While the extractor builds transactions directly from snapshot data, the generator infers synthetic transactions. In order to achieve this, it still needs access to the snapshot data. The reason for this is to get access to resource identifiers to be used in order to generate URLs. For instance, the generator creates queries, i.e., GET requests to query issues. If the respective resource names (i.e., issues ids) were generated randomly, almost all of those requests would fail with a 404. This is not very realistic: in practice, most GET requests would try to access existing resources and succeed. In order to model this, the generator needs to access the GHTorrent snapshot. 

\begin{figure}[!t]
	\centering
	\includegraphics[width=0.8\columnwidth]{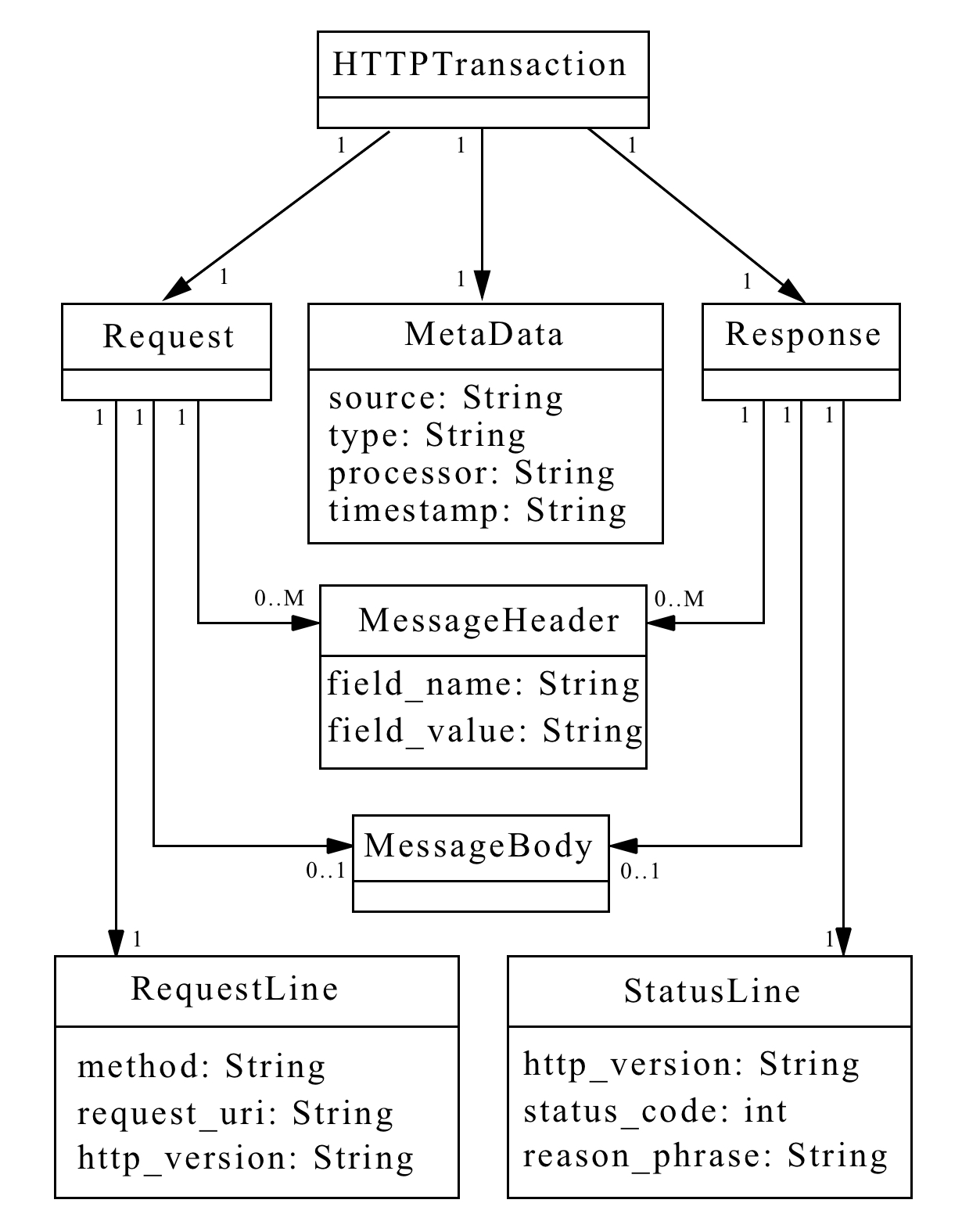}
	\caption{GHTraffic schema}
	\label{fig:objectschema}
\end{figure}

The transactions generated by both the extractor and the generator instantiate a simple model depicted in Figure~\ref{fig:objectschema}. This model is implemented in Java, i.e., each transaction has a transient in-memory representation as a Java object when the dataset is created. At the centre of this model are HTTP transactions, basically request/response pairs.

At the end of the pipeline is an \textit{Exporter} component that processes the transactions represented as Java objects and persists them by encoding/serialising using JSON. The structure of the JSON files produced is defined by JSON schemas~\cite{wright2016json}. Note that there are separate schemas for each HTTP method.

The implementation of the components discussed have some abstractions to facilitate alternative extraction, inference, and data representations. The overall processing model is lazy and stream-like, i.e., only a small number of records remain in memory at any time in order to make processing scalable. 

Processing can be customised by employing data filters (predicates). Only records matching certain criteria are processed. The main use case for this is filtering by URL and here in particular by the project. This allows us to build different editions of the dataset with certain target sizes. While there is a potentially easier way of doing this by just restricting the number of records being processed and included, using filters has an inherent advantage. GitHub data is fragmented by project and by filtering it accordingly, we are able to extract transactions that manipulate the same resources, reflecting the same issue being created and updated. This was, we can obtain \textit{coherent} subsets of the overall dataset that still reflect the service semantics derived from issue tracking workflows. 

\subsection{Extraction}
\begin{figure}[!t]
	\centering
	\includegraphics[width=\columnwidth]{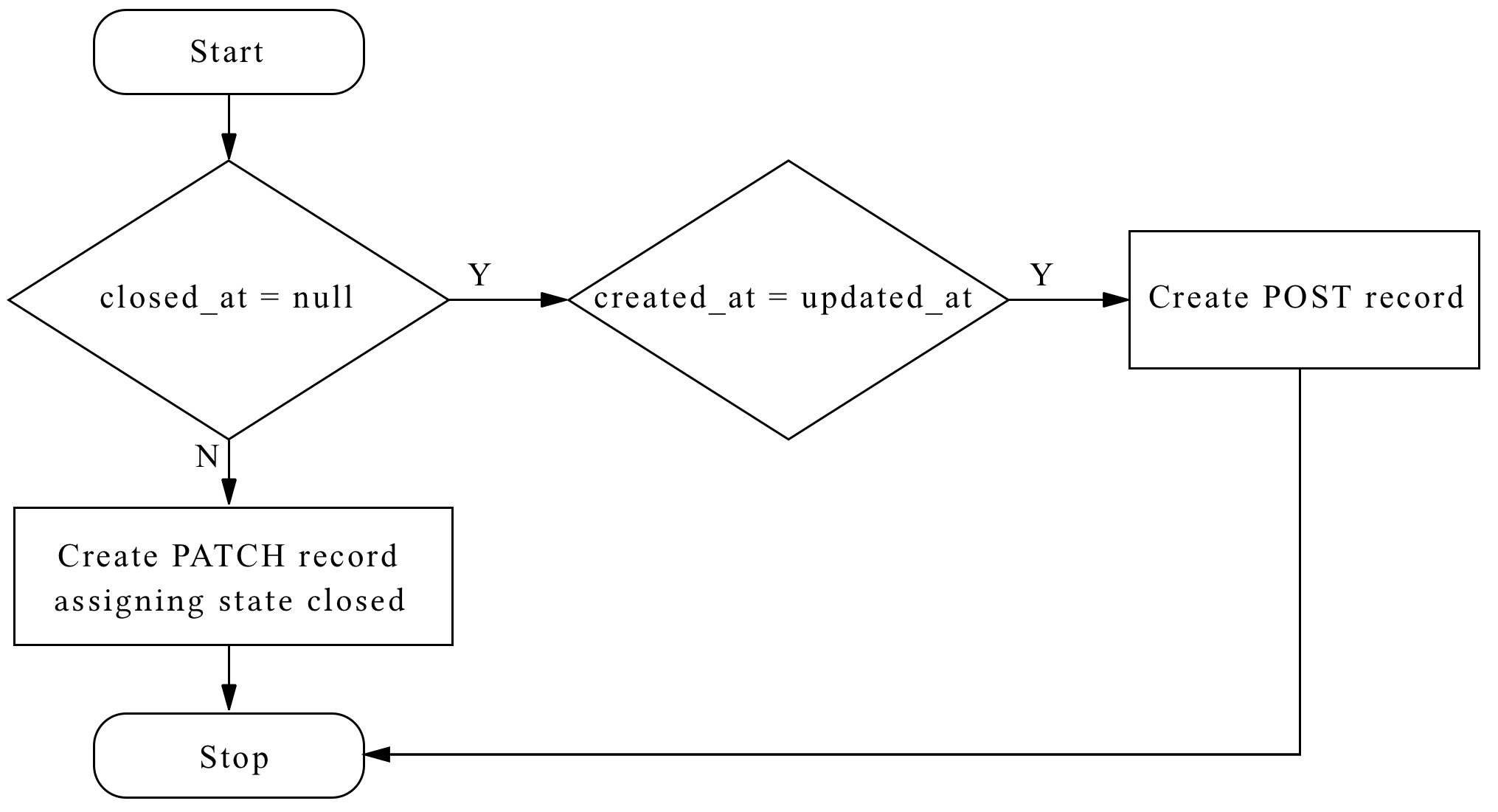}
	\caption{Extractor algorithm to process records}
	\label{fig:extractorflowchart}
\end{figure}

Each data record has  \texttt{{created\_at}},  \texttt{{updated\_at}}, and  \texttt{{closed\_at}} timestamps which enable us to trace lifecycle events of the issue. Using this data, the GHTraffic scripts produce transaction records. An overview of the process is shown in Figure~\ref{fig:extractorflowchart}.
For instance, a POST  transaction record is created in order to represent the creation of an issue at the time stipulated in the \texttt{{created\_at}} attribute. The value is converted to the standard date format used by the HTTP~\cite[Sect. 3.3]{fielding1999hypertext} and set as the value of \texttt{Date} header. 

Both the request and the response used headers as specified in the GitHub API documentation. This is a mix of standard HTTP headers and API-specific headers with names starting with ``\texttt{x-}''. In case the header values cannot be inferred from snapshot data, we use synthetic data. For example, we generate random token strings and use them as values for the \texttt{Authorization} headers. There is also a list of user agent strings to assign randomly as the value of the \texttt{User-Agent} headers. Further, for the request body, the script extracts the values of the \texttt{title, body, assignee, milestone, labels} parameters from the snapshot record and encodes it in JSON as stipulated in the API. The response creation process is analogous. Most of the values for the JSON-encoded response body are filled out with data directly taken from the snapshot. Besides, the GHTraffic script assigns the \texttt{{created\_at}} value to the \texttt{{updated\_at}} field. Further, it explicitly specifies \texttt{{closed\_at}: null}, \texttt{{closed\_by}: null}, \texttt{state: open} and \texttt{locked: false}. 

Every time a GitHub user updates an existing issue its \texttt{{updated\_at}} timestamp gets renewed with the date and time of the update. Marking an issue as closed is a special type of update, as an issue is not deleted, but its status is changed to \texttt{close}. In order to extract PATCH  transactions used to close issues, the script queries issues whose \texttt{{closed\_at}} value is not null and only those are processed by the extractor. The request and response messages are formed by following the GitHub Issues API documentation. Particularly, the \texttt{{closed\_at}} value is converted to the standard HTTP date format and set as the value of \texttt{Date} response header and the value of \texttt{{closed\_at}}, assign to the \texttt{{updated\_at}} field to set \texttt{{closed\_at}} and \texttt{{updated\_at}} columns' values same.  

Besides, an update might be a changing the title of an issue, changing its description, specifying users to assign the issue to, etc. However, we could not extract exactly what input data was used for editing an issue, therefore, we did not generate such transaction types.

\subsection{Synthesising Queries}
Only successful POST and PATCH transactions can be constructed by reverse-engineering the GHTorrent snapshot. In order to generate additional transaction records such as queries and delete requests, we had to resort to using synthetic data. The aim of generating synthetic data is to mimic transactions concerning several other HTTP request methods that are covered by the API and requests that fail, which indicated by an error HTTP status code. 

\begin{figure}[!t]
	\centering
	\includegraphics[width=0.35\textwidth]{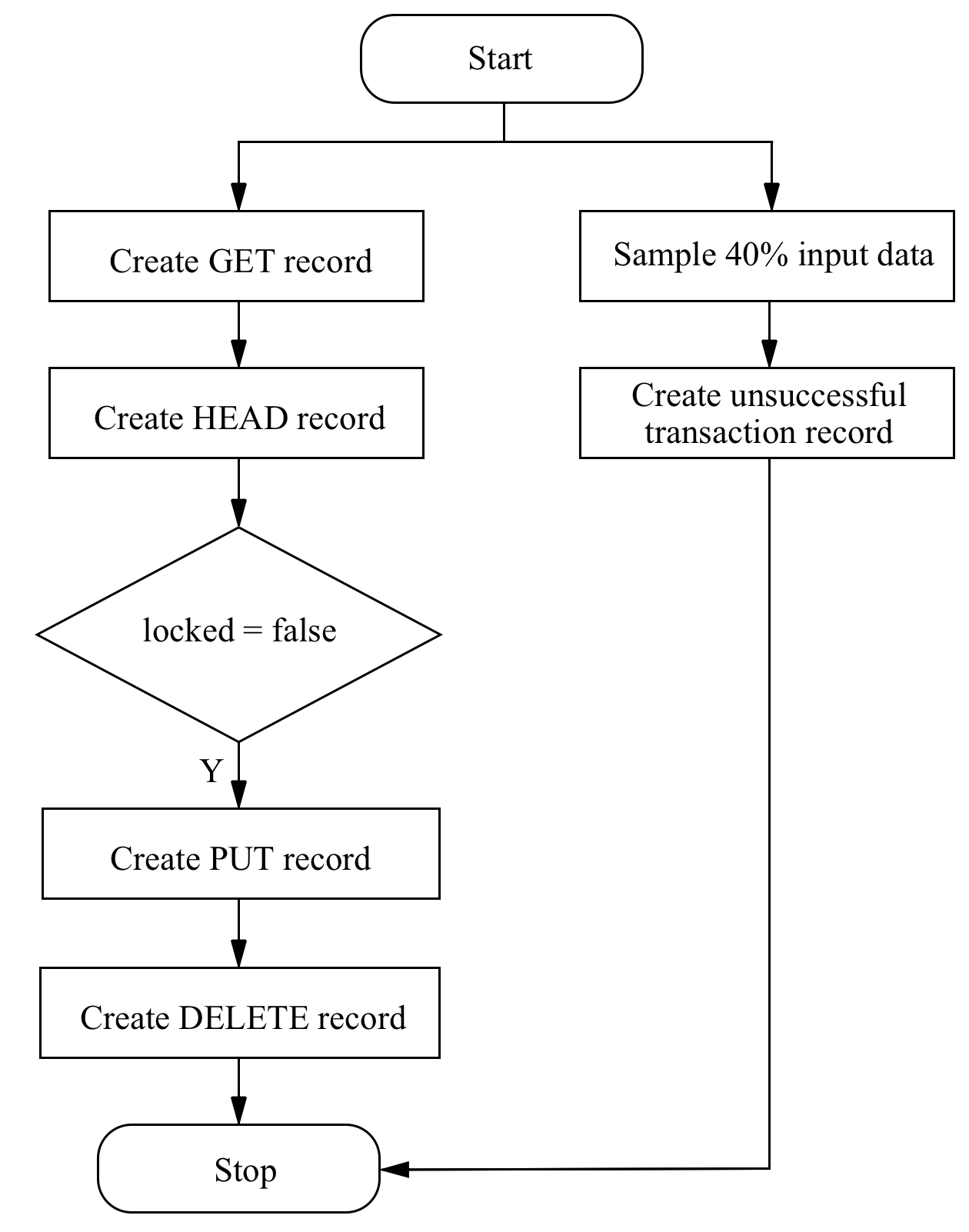}
	\caption{Algorithm to generate synthetic data}
	\label{fig:generatorflowchart}
\end{figure}

Figure~\ref{fig:generatorflowchart} shows the process of synthetic data generation.  
The script generated GET and HEAD transactions for each record in the snapshot. The process is analogous to the process described in Section~\ref{ssec:pipeline}. However, the \texttt{Date} response header is set to the system date and time at which the request is formed. Similarly, PUT transactions for locking an issue are generated for records with \texttt{locked} value set to \texttt{false} and followed by DELETE transactions for unlocking the respective issues using the format described in the GitHub API. 

Furthermore, the script produces unsuccessful transactions for all those HTTP methods by specifying requests:
\begin{itemize}
	\item without authorisation token
	\item with badly formatted URL
	\item without request body
	\item with invalidly formatted JSON body
\end{itemize}
All the respective transactions have an error status code as defined in the API and are generated from a sample of 40\% random records from the snapshot\footnote{The generator component needs to use at least 40\% of GHTorrent snapshot records in order to extract an adequate amount of unsuccessful transactions on particular projects}. More specifically, a message explaining the error is added to the response body as specified in the GitHub API. For this purpose, we performed experiments on a toy project repository for creating synthetic data that closely resemble real-world representation as we found that certain aspects of the GitHub Issue API are undocumented. Additionally, we generated a small number of GET requests that returned 500 status code, in order to represent system failures.

\subsection{Data Representation and Meta-Data}

The target format of the GHTraffic dataset is described by the UML class diagram as shown in Figure~\ref{fig:objectschema}. \texttt{HTTPTransaction} is the base element of the model. A transaction contains a single \texttt{Request} and \texttt{Response}. Each message could have any specific number of \texttt{MessageHeaders}. Additionally, a \texttt{MessageBody} is used to represent data associated with a request/response. \texttt{MetaData} is used to provide some additional information about the transaction record. The \texttt{source} attribute is set to \texttt{GHTorrent}, specifying the source of information. The \texttt{type} attribute is set to either \texttt{real-world} or \texttt{synthetic} depending on whether the data was directly derived from a GHTorrent record or synthesised as described above. The \texttt{processor} is the name of the script used to generate the record, i.e., this is the fully qualified name of a Java class. Finally, the \texttt{timestamp} field holds date and time when the record was created.

The actual JSON format of the dataset is defined by a set of JSON schemas for each transaction type (i.e., for each HTTP method). For space limitations, we do not include those schemas, but they can be found in the repository, in \textit{schemas} folder. These schemas comply with the JSON Schema draft 4 specification~\cite{galiegue2013json}. 

\section{Metrics} \label{sec:metrics}
The GHTraffic dataset comprises three different editions: Small (S), Medium (M), and Large (L). The S dataset includes HTTP transaction records created from \textit{google/guava}~\cite{googleguava} repository and takes up to 49.9 MB of disk space. Guava is a popular Java library containing utilities and data structures. It is a medium-sized large active project, and sourcing an edition from a single project has the advantage of creating a coherent dataset. The M dataset of size 345.2 MB includes records from the \textit{npm/npm}~\cite{npmnpm}. It is the popular de-facto standard package manager for JavaScript. The L dataset contains 3.73 GB of data that were created by selecting eight repositories containing large and active projects on GitHub as of 2015, including \textit{rails/rails}~\cite{rails}, \textit{docker/docker}~\cite{docker}, \textit{rust-lang/rust}~\cite{rust},  \textit{angular/angular.js}~\cite{angular}, \textit{twbs/bootstrap}~\cite{bootstrap}, \textit{kubernetes/kubernetes}~\cite{kubernetes},
\textit{Homebrew/homebrew}~\cite{homebrew}, and \textit{symfony/symfony}~\cite{symfony}.

Table~\ref{table:one},~\ref{table:two}, and~\ref{table:three} presents several metrics about the current status of these three datasets. 
\begin{table}[!t]
	\renewcommand{\arraystretch}{1.3}
	\caption{Transactions Per HTTP Method}
	\label{table:one}
	\centering
	\footnotesize
	\begin{tabular}{|c|c|c|c|}
			\hline
		Method & S & M & L \\	
			\hline
		POST   &7,193   &32,130  &508,664   \\
			\hline
		GET   &3,117   &22,692  &344,474   \\
			\hline
		PATCH   &4,286   &30,807  &468,080   \\
			\hline
		DELETE   &2,341   &16,457  &246,180   \\
			\hline
		PUT   &3,662   &15,945  &238,115   \\
			\hline
		HEAD   &1,796   &15,130  &245,127   \\
	\hline
	\end{tabular}
\end{table}

\begin{table}[!t]
	
	\renewcommand{\arraystretch}{1.3}
	\caption{Transactions Per HTTP Response Code}
	\label{table:two}
	\centering
	\footnotesize
	\begin{tabular}{|c|c|c|c|}
		\hline
		Response Code & S & M & L \\ 
		\hline
		200   &4,649   &22,163  &391,903   \\
		\hline
		201   &1,796   &8,808  &150,662   \\
		\hline
		204   &3,588   &5,756  &82,554   \\
		\hline
		400   &2,717   &13,807  &196,474   \\
		\hline
		401   &547   &19,302  &291,831   \\
		\hline
		404   &5,909   &43,658  &646,346   \\
		\hline
		422   &1,868   &12,626  &196,678   \\
		\hline
		500   &1,321   &7,041  &94,192   \\
		\hline
	\end{tabular}
\end{table}

\begin{table}[!t]
	
	\renewcommand{\arraystretch}{1.3}
	\caption{Transactions Per Record Type}
	\label{table:three}
	\centering
	\footnotesize
	\begin{tabular}{|c|c|c|c|}
		\hline
		Type & S & M & L \\ 
		\hline
		Real-world   &2,853   &13,355  &241,241   \\
		\hline
		Synthetic   &19,542   &119,806  &1,809,399   \\
		\hline
	\end{tabular}
\end{table}

\section{Accessing and Using GHTraffic} \label{sec:usage}
The different editions of the GHTraffic dataset can be downloaded by using the following URLs\footnote{The dataset is published on Zenodo~\cite{zenodo2016zenodo}. It is a data repository platform hosted at the European Organization for Nuclear Research Data Center, which was specifically designed to provide long-term preservation of all forms of research output.}: 
\begin{itemize} \small
	\item \url{https://zenodo.org/record/1034573/files/ghtraffic-S-1.0.0.zip}
	\item \url{https://zenodo.org/record/1034573/files/ghtraffic-M-1.0.0.zip}
	\item \url{https://zenodo.org/record/1034573/files/ghtraffic-L-1.0.0.zip}
\end{itemize}
We also provide access to the scripts used to generate GHTraffic, including a VirtualBox image with a pre-configured setup. Note that due to the use of random data generation these scripts will produce slightly different datasets at each execution. Using the scripts, users can modify the configuration properties in \textit{config.properties} file in order to create a customised version of GHTraffic dataset for their own use. The \textit{readme.md} file included in the distribution provides further information on how to build the code and run the scripts. Scripts can be accessed by cloning the repository \url{https://bitbucket.org/tbhagya/ghtraffic.git} or by downloading the pre-configured VirtualBox image from \url{https://zenodo.org/record/1034573/files/ghtraffic-artifact-1.0.0.zip}.

\section{Threats to Validity} \label{sec:threats}

As depicted in Table~\ref{table:three}, the size of synthetic data exceeds the size of data extracted from the snapshot by a factor of nine. This leaves the possibility that GHTraffic does not reflect realistic workloads. To mitigate this threat, we ensured that the request/response format for all these transaction types was sampled and validated using a toy GitHub repository. While this does not mean that these transactions have actually occurred, it guarantees that they are syntactically and semantically correct. The representation of the transactions has information about whether they are synthetic or not, and users of GHTraffic can use this to completely remove or reduce the ratio of synthetic data by applying filters.

We acknowledge that GHTraffic was generated from a 2-year-old snapshot of GHTorrent.  As noted earlier, this design decision was made to produce a dataset large enough to facilitate the use cases described, but still manageable with typical resources available to researchers and practitioners. We also provide access to the scripts used to generate GHTraffic, and users can utilise these scripts in order to generate customised versions from newer instances of GHTorrent if needed.

\section{Conclusion} \label{sec:conclusion}
In this paper, we have described the GHTraffic dataset suitable for experimenting on various aspects of service-oriented computing. It is derived from reverse-engineering a GHTorrent snapshot according to the GitHub Issue API specification. We hope that this dataset will find uses in many areas of research. 

In future work, it would be interesting to extend this by adding similar datasets from other service providers, using similar processes and tools.

% references
\def\IEEEbibitemsep{0pt plus .5pt}
\bibliographystyle{IEEEtran}
\bibliography{IEEEabrv,bibliography}

\end{document}